\documentclass[a4paper,11pt]{article}
\pdfoutput=1 % if your are submitting a pdflatex (i.e. if you have
\usepackage{jcappub} % for details on the use of the package, please
\usepackage{mathabx}
\usepackage{mathtools}

\usepackage[T1]{fontenc} % if needed

\title{\boldmath Updated detection prospects for relic neutrinos using coherent scattering}

\author[]{Jack D. Shergold}

\affiliation[]{Institute for Particle Physics Phenomenology, Department of Physics,\\Durham University,\\Durham, UK}

\emailAdd{jack.d.shergold@durham.ac.uk}

\abstract{We review the existing proposals to detect relic neutrinos using the coherent scattering of a neutrino wind on a test mass. By considering the transformation of the neutrino momentum between reference frames, we demonstrate that the induced acceleration scales with the square of the neutrino mass for unclustered neutrinos, contrary to the existing literature. In addition, we show that there is a large contribution to this effect from coherent neutrino-electron scattering, which can exceed the neutrino-nucleus component by nearly an order of magnitude. Unfortunately, we find that even with this enhancement there are no existing experiments or proposals capable of detecting relic neutrinos using this method.}

\begin{document}
\maketitle
\flushbottom
\section{Introduction}
When neutrinos were first predicted by Pauli in 1930 and later incorporated into a theory of $\beta$-decay~\cite{Fermi:1933jpa}, it was expected that they would never be observed as a consequence of their feebly interacting nature. The groundbreaking experiment of Cowan and Reines defied this expectation in 1953 and detected the electron neutrino using inverse $\beta$-decay~\cite{Reines:1953pu,Cowan:1956rrn}. Since then, several experiments have gone on to detect neutrinos originating from a range of both terrestrial and astrophysical sources~\cite{Danby:1962nd, Kamiokande-II:1987idp, Super-Kamiokande:1998kpq, SNO:2001kpb, DONUT:2000fbd, T2K:2011ypd, IceCube:2013cdw, KamLAND:2002uet, STEREO:2018rfh, MiniBooNE:2010idf, Davis:1968cp, GALLEX:1992gcp, MINOS:2006foh, Borexino:2008dzn}. In doing so, these experiments have revealed two additional neutrino species~\cite{Danby:1962nd, DONUT:2000fbd}, neutrino flavour oscillations and consequently non-zero neutrino masses~\cite{Super-Kamiokande:1998kpq, SNO:2001kpb, T2K:2011ypd, KamLAND:2002uet, MINOS:2006foh}, and more recently evidence for CP-violation in the lepton sector~\cite{T2K:2019bcf}.  

Despite the remarkable progress made by neutrino experiments, relic neutrinos from the early universe have thus far eluded detection due to their low energy and tiny interaction cross sections. Several proposals have been put forward to detect the cosmic neutrino background (C$\nu$B). The most well-known of these was proposed by Weinberg in 1962~\cite{Weinberg:1962zza}, which aims to capture neutrinos on radioactive nuclei. This principle is currently being developed into the PTOLEMY experiment~\cite{PTOLEMY:2018jst}. Other proposals include scanning the high energy cosmic ray neutrino flux for an absorption line at the $Z$-resonance~\cite{Eberle:2004ua}; using an ion storage ring to resonantly capture neutrinos~\cite{Bauer:2021uyj}; searching for modifications in atomic de-excitation spectra due to the Pauli exclusion principle~\cite{Yoshimura:2014hfa}; as well as using either a torsion balance or laser interferometer to observe tiny accelerations induced by a relic neutrino wind~\cite{Duda:2001hd, Domcke:2017aqj}. The last of these can be decomposed into two effects: the first utilises a neutrino-antineutrino asymmetry (or left-right helicity asymmetry, for Majorana neutrinos) induced torque on a magnet, commonly known as the Stodolsky effect~\cite{Stodolsky:1974aq,Langacker:1982ih}. The second is due to momentum transfer by the coherent scattering of neutrinos.

In this paper we will focus on the second effect, which is considerably enhanced by their macroscopic de Broglie wavelength at low energies. As we will show, there is a significant contribution to this acceleration due to electron-neutrino scattering that has not been explored in detail by previous works. In addition, we will find a different scaling for the acceleration with neutrino mass at intermediate neutrino temperatures, and show that there exists a temperature independent upper limit to the acceleration for sufficiently massive neutrinos. 

The detection of relic neutrinos would provide a window into big bang nucleosynthesis. Additionally, their low energy makes any detection attempt naturally sensitive to the neutrino mass, the upper bound on which currently stands at $m_\nu < 0.8\,\mathrm{eV}$ set by KATRIN~\cite{Aker:2021gma}. On the contrary, a null result would be indicative of a modified thermal history of the universe, calling both the $\Lambda$CDM model and Standard Model of particle physics (SM) into question.

The remainder of this paper will be structured as follows. In Section~\ref{sec:cnb} we will briefly review the evolution of the C$\nu$B in the standard scenario, before deriving the acceleration of a test mass in the presence of a neutrino wind due to coherent scattering in Section~\ref{sec:gfsq}. We will discuss the expected performance of an experiment utilising this effect in Section~\ref{sec:exp}, before concluding in Section~\ref{sec:conc}. 
%
%------------------------------------------------------------------------
%
\section{Neutrino thermal history}\label{sec:cnb}
In the early universe, neutrinos remain in thermal equilibrium with the SM thermal bath through weak scattering on electrons, with scattering rate $\Gamma_{\nu-e} \sim G_F^2 T_\nu^5$,
where $G_F$ is Fermi's constant and $T_\nu$ is the neutrino temperature\footnote{We work in units with $\hbar = c = k_B = 1$ throughout.}. As the universe expands, the SM thermal bath cools and neutrinos decouple when the mean scattering rate equals the expansion rate $H \sim \sqrt{G_N} T_\nu^2$, with $G_N$ the gravitational constant. This corresponds to a decoupling temperature $T_\mathrm{dec} \simeq 1\,\mathrm{MeV}$, at which point the time between neutrino-electron scattering events is approximately one age of the universe. At around the same temperature, the process $\gamma \to e^+ e^-$ freezes out and free electron-positron pairs annihilate into photons. As entropy must be conserved, this reheats the photon bath to a temperature $T_\gamma = (11/4)^{1/3}\, T_\nu$, now out of equilibrium with the C$\nu$B. Based on measurements of the present-day cosmic microwave background (CMB) temperature~\cite{Fixsen:2009ug}, this places the C$\nu$B temperature at $T_{\nu,0} = 0.168\,\mathrm{meV}$. 

We further note that although relic neutrinos were produced as weak eigenstates, they have long since decohered and exist today as freely propagating mass eigenstates~\cite{Long:2014zva}. Whilst there is no lower bound on the mass of lightest neutrino mass eigenstate, the squared mass splittings measured by neutrino oscillation experiments tell us that $m_{\nu_2} \gtrsim 8.6\,\mathrm{meV}$, $m_{\nu_3} \gtrsim 50.2\,\mathrm{meV}$ in the normal mass hierarchy (NH), and $m_{\nu_1} \gtrsim 50.0\,\mathrm{meV}$, $m_{\nu_2} \gtrsim 50.7\,\mathrm{meV}$ in the inverted mass hierarchy (IH)~\cite{Esteban:2020cvm}. As a result, at least two of the three neutrino states will always be non-relativistic in the standard scenario with $T_\nu = T_{\nu,0}$, which as we will see in Section~\ref{sec:gfsq} has a significant effect on the induced accelerations.

If, unlike cold dark matter, neutrinos do not cluster in our galaxy due to their low mass, it is reasonable to suggest that their reference frame should coincide with that of the CMB. In this instance the neutrino wind will be generated by the Earth's velocity relative to the CMB reference frame, which from CMB dipole measurements is 
$\beta_\Earth^{\mathrm{CMB}} \simeq 10^{-3}$~\cite{Amendola:2010ty,Ferreira:2020aqa}. Within the CMB frame, unclustered neutrinos are expected to follow their equilibrium distribution, redshifted~\cite{Gnedin:1997vn}
\begin{equation}
    f_\nu(p_\nu) = \frac{1}{\exp{(p_\nu/T_\nu)}+1}, 
\end{equation}
which importantly is independent of their mass. This in turn yields a mass independent number density $n_{\nu,0} \simeq 56\,\mathrm{cm}^{-3}$ per degree of freedom for $T_{\nu} = T_{\nu,0}$, as well as a mean neutrino momentum $\bar p_\nu \simeq 3.15\, T_\nu$. 

We contrast this to the case where neutrinos are clustered, where instead the C$\nu$B reference frame is that of the Milky Way. In this scenario, the relative motion of the Earth to the C$\nu$B is $\beta_\Earth^{\mathrm{MW}}\simeq 7.6\cdot 10^{-4}$~\cite{Vera:2020}. Neutrinos are only able to cluster efficiently if their mean velocity $\bar\beta_\nu$ does not exceed the escape velocity the galaxy $\beta_\mathrm{esc} \simeq 1.8\cdot 10^{-3}$~\cite{Kafle:2014xfa}. By setting $\bar\beta_\nu = \beta_{\mathrm{esc}}$ we find the temperature below which neutrinos are able to cluster, $T_{\nu,\mathrm{cluster}}\simeq 5.8\cdot 10^{-4}\,m_\nu$. In the standard thermal history, neutrinos are therefore only able to cluster with masses $m_\nu \gtrsim 0.29\,\mathrm{eV}$, which are not yet ruled out by direct detection experiments. Relic neutrino clustering is thus a realistic possibility which could lead to a local neutrino number density per mass eigenstate $n_{\nu_i} > n_{\nu,0}$, whilst also modifying their momentum profile.

Of course, modified number densities are also possible in the case of unclustered relic neutrinos. For example, adding just one bosonic degree of freedom with late decays to photons sets the C$\nu$B temperature $T_\nu = (11/12)^{1/3}\, T_{\nu,0}$, reducing the number density to $n_{\nu_i} = (11/12)\, n_{\nu,0}$. We will therefore leave the relic neutrino density parameter $f_{c,i} = n_{\nu_i} / n_{\nu,0}$ as a free parameter throughout the remainder of this paper, which may differ for each neutrino degree of freedom.
%
%----------------------------------------------------------------------
%
\section{Coherent neutrino scattering}\label{sec:gfsq}
We now calculate the acceleration of a test mass due to momentum transfer by the neutral scattering of relic neutrinos. This idea has already been discussed by several authors~\cite{Opher:1974drq,Gelmini:2004hg,Vogel:2015vfa,Shvartsman:1982sn,Ringwald:2004np,Duda:2001hd,Domcke:2017aqj,Cabibbo:1982bb, Smith:1983jj, Lewis:1979mu}, however, there remains some disagreement in the scaling of this effect with the relic neutrino mass and temperature which we will attempt to resolve here. 
We will work in the flavour basis, noting that the weak eigenstate masses and number densities are related to those in the mass basis by $m_{\nu_\alpha} = \sum_i |U_{\alpha i}|^2 \,m_{\nu_i}$ and $n_{\nu_\alpha} = \sum_i |U_{\alpha i}|^2 \,n_{\nu_i}$, where $U_{\alpha i}$ is an element of the PMNS matrix and $\alpha \in \{e,\mu,\tau\}$. To avoid cluttered notation we will first derive the acceleration for a single weak eigenstate and then restore the subscripts when we sum over all states. 

The neutrino-nucleus scattering induced acceleration of the test mass with mass $M$ is
\begin{equation}\label{eq:accNN}
    a^N = \frac{\Gamma_{\nu-N}\, \Delta p_\nu}{M},
\end{equation}
where $\Gamma_{\nu-N}$ is the neutrino-nucleus scattering rate and $\Delta p_\nu$ is the average momentum transfer by a single scattering event. We can express the scattering rate in terms of the coherent neutrino-nucleus scattering cross section $\sigma_{\nu-N}$ as $\Gamma_{\nu-N} = N \phi_\nu\, \sigma_{\nu-N}$, for a test mass comprised of $N$ nuclei in a background of relic neutrinos with flux $\phi_\nu = (p_\nu/E_\nu)\, n_{\nu}$. The coherent neutrino-nucleus scattering cross section for Dirac neutrinos is given by~\cite{Domcke:2017aqj,Formaggio:2012cpf}
\begin{equation}
     \sigma_{\nu-N} \simeq \frac{G_F^2}{4\pi} (A-Z)^2 E_\nu^2,
\end{equation}
for a nucleus with $A$ nucleons and $Z$ protons, where $E_\nu$ is the mean neutrino energy in the rest frame of the target. Combining everything together so far, we arrive at 
\begin{equation}\label{eq:nnxsec}
    a^N = \frac{G_F^2}{4\pi} \frac{N_A}{m_A} \frac{(A-Z)^2}{A} E_\nu\,p_\nu\,\Delta p_\nu \,n_\nu,
\end{equation}
where $N_A$ is Avogadro's number and we have introduced the "Avogadro mass", $m_A = 1\,\mathrm{g}\,\mathrm{mol}^{-1}$.

If relic neutrinos are non-relativistic, they will have a macroscopic de Broglie wavelength $\lambda_\nu = 2\pi/p_\nu$. This leads to a significant enhancement in the induced accelerations due to coherence effects~\cite{Opher:1974drq, Shvartsman:1982sn, Smith:1983jj, Lewis:1979mu, Domcke:2017aqj, Duda:2001hd}, proportional to the number of target nuclei within a volume $\lambda_\nu^3$
\begin{equation}
    N_c = \left(\frac{2\pi}{p_\nu}\right)^3 \frac{N_A}{A\,m_A}\rho,
\end{equation}
where $\rho$ denotes the mass density of the target. In order to maximise the coherence effects the target should be chosen such that its spatial extent is of order $\lambda_\nu$, else coherence will be lost due to destructive interference~\cite{Shvartsman:1982sn}. Including the coherence effects, the acceleration due to the neutrino wind is
\begin{equation}\label{eq:accSingle}
    a_c^N = N_c\, a^N = 2\pi^2 G_F^2 \left(\frac{N_A}{m_A}\frac{A-Z}{A}\right)^2 \frac{E_\nu}{p_\nu^2}\,\Delta p_\nu\, n_\nu\, \rho.
\end{equation}
As it must be true that $p_\nu \geq \Delta {p_\nu}$,~\eqref{eq:accSingle} naturally favours scenarios in which neutrinos have lower momentum, as expected by the introduction of the coherence factor.

If we had instead considered Majorana neutrinos, the absence of vector currents suppresses the cross section~\eqref{eq:nnxsec} by a factor of $(p_\nu/E_\nu)^2$~\cite{Duda:2001hd}, which is $\ll 1$ for non-relativistic neutrinos. Thus whilst it is possible to detect Majorana neutrinos using this method, it is considerably more challenging than detecting Dirac neutrinos. For that reason, we will limit our discussion to Dirac neutrinos for the remainder of this paper. 

\subsection{Estimating the average momentum transfer}
Next, we calculate the average momentum transfer from relic neutrinos to the target, $\Delta p_\nu$. If the Earth were stationary relative the C$\nu$B, we would expect that $\Delta p_\nu = 0$ on account of an equal number of neutrinos with equal and opposite momenta striking our target. On the contrary, the relative motion of the Earth with respect to the C$\nu$B blueshifts those neutrinos in the path of the Earth whilst redshifting those in its wake. This in turn leads to a non-zero momentum transfer that scales proportional to the Earth's velocity $\beta_\Earth$ relative to the C$\nu$B.

\begin{figure}[tbp]
\centering 
\includegraphics[width=.45\textwidth]{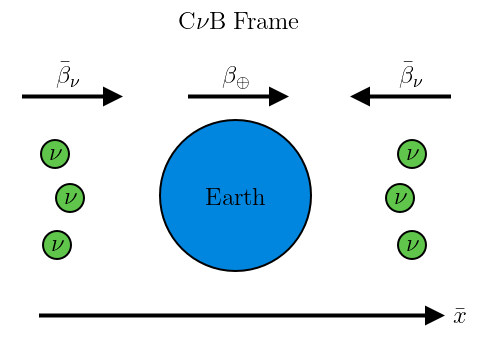}
\hfill
\includegraphics[width=.45\textwidth]{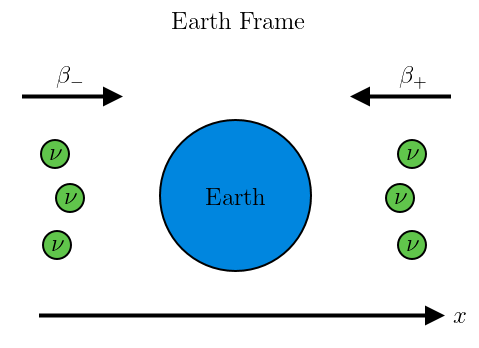}
\caption{\label{fig:frames} Setup used to estimate the average momentum transfer, $\Delta p_\nu$. Left: The Earth moves relative to the C$\nu$B frame with velocity $\beta_\Earth$, within which neutrinos move with mean velocity $\bar \beta_\nu$. Right: In the Earth's reference frame, neutrinos move with velocities $\beta_+$ or $\beta_-$.}
\end{figure}

To make an estimate of this effect, we consider a simple 1-dimensional setup with the Earth travelling along the positive $x$-axis with velocity $\beta_\Earth$ in the C$\nu$B frame, in which the neutrinos can either be left or right travelling with velocity $\bar\beta_\nu$. We sketch this setup in Figure~\ref{fig:frames}. The velocities of the neutrinos in the rest frame of the Earth will then be given by
\begin{equation}\label{eq:velocityAddition}
    \beta_\pm = \frac{\bar\beta_\nu \pm \beta_\Earth}{1\pm \bar\beta_\nu \beta_\Earth},
\end{equation}
where the $+$ $(-)$ denotes the left (right) travelling neutrinos in the C$\nu$B frame. These velocities can be related to the Earth rest frame neutrino momenta and energy by
\begin{equation}\label{eq:ppm}
    p_\pm = \beta_\pm E_\pm, \quad E_\pm = \sqrt{p_\pm^2 + m_\nu^2}.
\end{equation}
Supposing then that the momentum transfer by a single neutrino is of order its momentum, the average momentum transferred to the test mass by each scattering event will be
\begin{equation}\label{eq:momTrans}
    \Delta p_\nu = \frac{\beta_+ p_+ - \beta_- p_-}{\beta_+ + \beta_-} \simeq  \frac{2-\bar\beta_\nu^2}{\sqrt{1-\bar\beta_\nu^2}}\,\beta_\Earth\,m_\nu+ \mathcal{O}(\beta_\Earth^2),
\end{equation}
where the factors of $\beta_\pm$ appearing after the first equality account for the fact that the Earth will encounter a larger flux of neutrinos that are left travelling than right travelling.

Before continuing, we make some important comments about this result. Whilst it is true that the momentum of C$\nu$B neutrinos is independent of their mass, the average momentum transferred to the test mass is not necessarily. The origin of this effect is in~\eqref{eq:velocityAddition}; the relative motion of the Earth to the C$\nu$B induces a mass-independent shift in the Earth rest-frame neutrino velocity. The subsequent change in momentum then depends on the neutrino mass through~\eqref{eq:ppm}, provided that relic neutrinos are non-relativistic. 

We now turn our attention to $\beta_\Earth$ and $\bar\beta_\nu$, considering both clustered and unclustered relic neutrino scenarios. If neutrinos are clustered then we expect that $\beta_\Earth = \beta_\Earth^{\mathrm{MW}}$, corresponding to the Earth's velocity about the galactic centre. In this case, the velocity dispersion of relic neutrinos should be of the same order as the velocity dispersion of the galaxy, and so we set $\bar\beta_\nu = \beta_\Earth^{\mathrm{MW}}$. This yields for clustered relic neutrinos,
\begin{equation}
    \Delta p_\nu \simeq 2m_\nu \beta_\Earth^{\mathrm{MW}}.
\end{equation}
Alternatively, if relic neutrinos are unclustered, then their reference frame should coincide with that of the CMB and we set $\beta_\Earth = \beta_\Earth^{\mathrm{CMB}}$. Within the CMB frame,  neutrinos move with velocities determined by their temperatures, $\bar\beta_\nu = \bar p_\nu/\bar E_\nu$, where $\bar E_\nu = \sqrt{\bar p_\nu^2 + m_\nu^2}$ is the mean energy of neutrinos in the C$\nu$B frame. Taking the appropriate limits, the average momentum transfer for unclustered neutrinos is given by
\begin{equation}\label{eq:momTransLims}
    \Delta p_\nu \simeq 
    \begin{cases}
        3.15 \,T_\nu \beta_\Earth^{\mathrm{CMB}},\quad&T_\nu\gg m_\nu, \\
        2m_\nu\beta_\Earth^{\mathrm{CMB}},\quad& T_\nu \ll m_\nu.
    \end{cases}
\end{equation}
Thus we recover the dipole effect $\Delta p_\nu \,\propto\, \beta_\Earth$ observed in existing literature, but find a different scaling relation for non-relativistic, non-clustered relic neutrinos~\cite{Duda:2001hd,Domcke:2017aqj}. For completeness, we note that the result~\eqref{eq:momTrans} is only valid for massive neutrinos. However, applying the same procedure using Doppler shift for massless neutrinos recovers the relativistic limit of~\eqref{eq:momTransLims}.

\subsection{Non-relativistic vs highly non-relativistic neutrinos}
To proceed further, we must be careful when dealing with $E_\nu$ and $p_\nu$, which are the mean energy and momentum of relic neutrinos in the Earth's rest frame. Intuitively, the distinction between Earth rest frame and C$\nu$B frame quantities is only relevant for non-relativistic neutrinos, as the Earth's velocity is much smaller than that of a relativistic neutrino. Additionally, we note that since $\beta_\Earth \ll 1$, the transformation between frames will not convert a non-relativistic neutrino into a relativistic one. Thus we expect that $E_\nu \simeq \bar E_\nu \simeq m_\nu$ for all non-relativistic neutrinos in the C$\nu$B frame.

However, it is not always the case that $p_\nu = \bar p_\nu$. Sufficiently cold unclustered neutrinos will have a C$\nu$B frame velocity $\bar\beta_\nu \ll \beta_\Earth^{\mathrm{CMB}}$, such that their Earth reference frame velocity is dominated by the relative motion of the Earth. On the other hand, the Earth frame velocity of warm neutrinos will be dominated by their temperature. We therefore choose the root-mean-square (rms) value of $\beta_+$ and $\beta_-$ for the Earth frame neutrino velocity, giving an Earth frame momentum for unclustered neutrinos
\begin{equation}\label{eq:earthMomentum}
    p_\nu = m_\nu\, \sqrt{\frac{\beta_+^2 + \beta_-^2}{2-\beta_+^2 -\beta_-^2}} \simeq \begin{cases}
        3.15\, T_\nu, \quad& T_\nu \gg m_\nu, \\
        m_\nu \bar \beta_\nu, \quad& T_{\nu,\mathrm{cold}} \ll T_\nu \ll m_\nu, \\
        m_\nu \beta_\Earth^{\mathrm{CMB}}, \quad& T_\nu \ll T_{\nu,\mathrm{cold}}.
    \end{cases}
\end{equation}
By setting $\bar\beta_\nu = \beta_\Earth^{\mathrm{CMB}}$, we find that the transition between the two cases occurs at a temperature $T_{\nu,\mathrm{cold}} \simeq 3.2\cdot 10^{-4} \,m_\nu$. This is below the clustering temperature $T_{\nu,\mathrm{cluster}}$ for all values of the neutrino mass, and so we might expect this effect to be irrelevant. However, it is shown in~\cite{LoVerde:2013lta} that the fraction of clustered neutrinos is small in galaxies whose dark matter haloes have comparable mass to that of the Milky Way, around $10^{12}$ solar masses~\cite{SDSS:2008nmx}. In particular, for neutrinos with masses of order the clustering mass, $m_\nu \simeq 0.3\,\mathrm{eV}$, the expected fraction of clustered neutrinos is $\mathcal{O}(< 10^{-2})$.
%However, it is in shown in~\cite{LoVerde:2013lta} that the fraction of clustered neutrinos with masses of order the clustering mass, $m_\nu \simeq 0.3\,\mathrm{eV}$, is $\mathcal{O}(<10^{-2})$ for galaxies with dark matter halo masses comparable to that of the Milky Way, around $10^{12}$ solar masses~\cite{SDSS:2008nmx}.
As a result, we will treat clustered and unclustered neutrinos separately below $T_{\nu,\mathrm{cluster}}$. To remain consistent, we also use~\eqref{eq:earthMomentum} for clustered neutrinos, which yields
\begin{equation}
    p_\nu \simeq \sqrt{2}\, m_\nu \beta_\Earth^{\mathrm{MW}}.
\end{equation}
Summing over all states and assuming the same number densities for neutrinos and antineutrinos, the total acceleration due to the neutrino wind is
\begin{equation}\label{eq:acctot}
    a_{c,\mathrm{tot}}^N\simeq 4\pi^2 G_F^2 \left(\frac{N_A}{m_A}\frac{A-Z}{A}\right)^2\! \rho \sum_\alpha {n_{\nu_\alpha}} \times 
    \begin{dcases}
        \beta_\Earth^{\mathrm{CMB}}, \! & T_{\nu_\alpha} \gg m_{\nu_\alpha}, \\
        \left(\frac{\sqrt{2}\,m_{\nu_\alpha}}{3.15\,T_{\nu_\alpha}}\right)^2\beta_\Earth^{\mathrm{CMB}}, \! & T_{\nu,\mathrm{cold}} \ll T_{\nu_\alpha} \ll m_{\nu_\alpha}, \\
        \frac{2}{\beta_\Earth^{\mathrm{CMB}}}, \! & T_{\nu_\alpha} \ll T_{\nu,\mathrm{cold}},\\
        \frac{1}{\beta_\Earth^{\mathrm{MW}}}, \! & \mathrm{clustered},
    \end{dcases}
\end{equation}
where $T_{\nu_\alpha}$ is the temperature of neutrino flavour eigenstate $\alpha$. Thus we find an effect that scales like $(m_\nu/T_\nu)^2$ for unclustered neutrinos at intermediate temperatures, in contrast to the existing literature where the effect scales like $(m_\nu/T_\nu)$ due to a mass independent choice of $\Delta p_\nu$. The result~\eqref{eq:acctot} also demonstrates that there is a mass independent upper limit to the acceleration for sufficiently cold unclustered neutrinos, such that the best and worse case scenarios differ by at most $(\beta_\Earth^{\mathrm{CMB}})^2 \simeq 10^{-6}$ for the same neutrino density. 

It is instructive to get a numerical estimate of the size of this effect. To do so we consider a silicon target with mass density $\rho \simeq 2.33\,\mathrm{g}\,\mathrm{cm}^{-3}$, similar to the mirrors which might be found at a laser interferometry experiment. This gives
\begin{equation}
    \frac{a_{c,\mathrm{tot}}^N}{1\,\mathrm{cm}\,\mathrm{s}^{-2}}\simeq \sum_\alpha f_{c,\alpha}\times \begin{dcases}
        3.05\cdot 10^{-34}, \quad& T_{\nu_\alpha} \gg m_{\nu_\alpha}, \\
        6.16\cdot 10^{-35}\left(\frac{m_{\nu_\alpha}}{T_{\nu_\alpha}}\right)^2, \quad& T_{\nu,\mathrm{cold}} \ll T_{\nu_\alpha} \ll m_{\nu_\alpha}, \\
        6.10\cdot 10^{-28}, \quad& T_{\nu_\alpha} \ll T_{\nu,\mathrm{cold}},\\
        4.02\cdot 10^{-28}, \quad& \mathrm{clustered},
    \end{dcases}
\end{equation}
where we have introduced $f_{c,\alpha} = \sum_i |U_{\alpha i}|^2\, f_{c,i}$. This gives an acceleration $a^N_{c,\mathrm{tot}} \gtrsim 10^{-30}\,\mathrm{cm}\,\mathrm{s}^{-2}$ in the standard scenario where at least two neutrinos are non-relativistic.

Assuming that $f_{c,\alpha}$ has no temperature dependence, the situation becomes worse for hotter neutrino backgrounds. However, it is more reasonable to suggest that $n_{\nu_\alpha}$ scales with the equilibrium number density, in which case the overdensity factor $f_{c,\alpha} = (T_{\nu_\alpha}/T_{\nu,0})^3$ for unclustered neutrinos and hotter backgrounds provide larger accelerations. The situation becomes somewhat more complicated for clustered neutrinos. Colder neutrinos are able to cluster more efficiently, which may lead to larger overdensity factors despite the reduced equilibrium number density. For simplicity, we will only consider the two cases $f_{c,\alpha} = 1$ and $f_{c,\alpha} = (T_{\nu_\alpha}/T_{\nu,0})^3$ in what follows. 
\subsection{Neutrino-electron scattering}
In addition to the neutrino-nucleus scattering discussed so far, there may be a contribution to the acceleration from neutrino-electron scattering, which was identified in~\cite{Domcke:2017aqj} but not explored in detail. Unlike neutrino-nucleus scattering, the cross section is sensitive to the flavour composition of the C$\nu$B~\cite{Marciano:2003eq},
\begin{equation}
    \sigma_{\nu_{\alpha}-e} = \frac{7G_F^2}{4\pi} k_\alpha E_\nu^2,
\end{equation}
for neutrino energies much less than the electron mass, where $k_\alpha = 1$ for electron neutrinos and $k_\alpha = 3/7$ for muon and tau neutrinos. This is much smaller than the neutrino-nucleus scattering cross section~\eqref{eq:nnxsec} for heavy nuclei due to the absence of the $(A-Z)^2$ term. However, for every nucleus in the target there are $Z$ electrons, enhancing both the collision rate and the coherence factor $N_c$ by a factor of $Z$ relative to the neutrino-nucleus case. We therefore find an acceleration due to neutrino-electron scattering
\begin{equation}
    a_c^e = 7k_\alpha\left(\frac{Z}{A-Z}\right)^2 a_c^N, 
\end{equation}
which is considerably larger than acceleration due to neutrino-nucleus scattering for the silicon target discussed previously. 

\begin{figure}[tbp]
\centering 
\includegraphics[width=.49\textwidth]{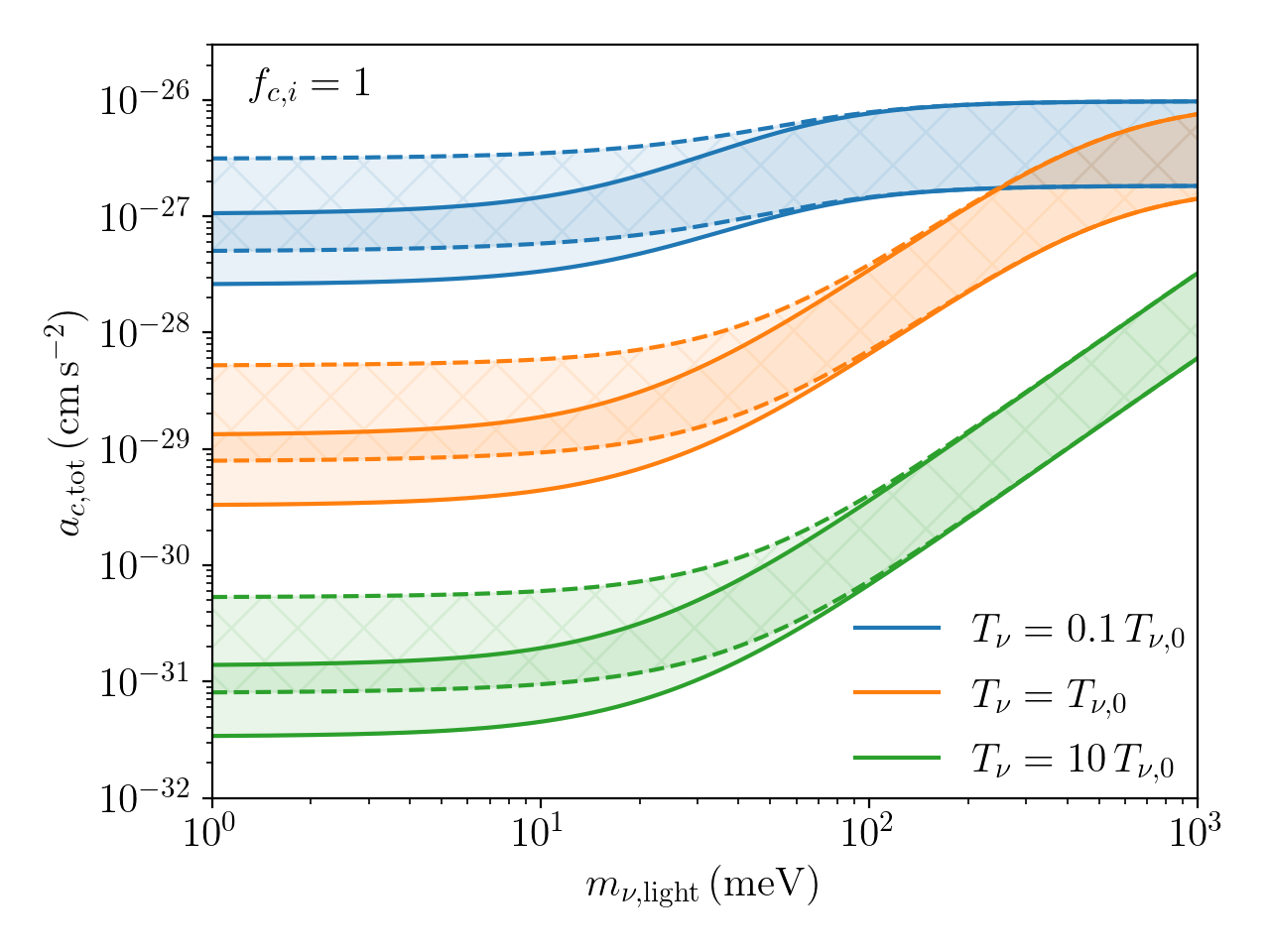}
\hfill
\includegraphics[width=.49\textwidth]{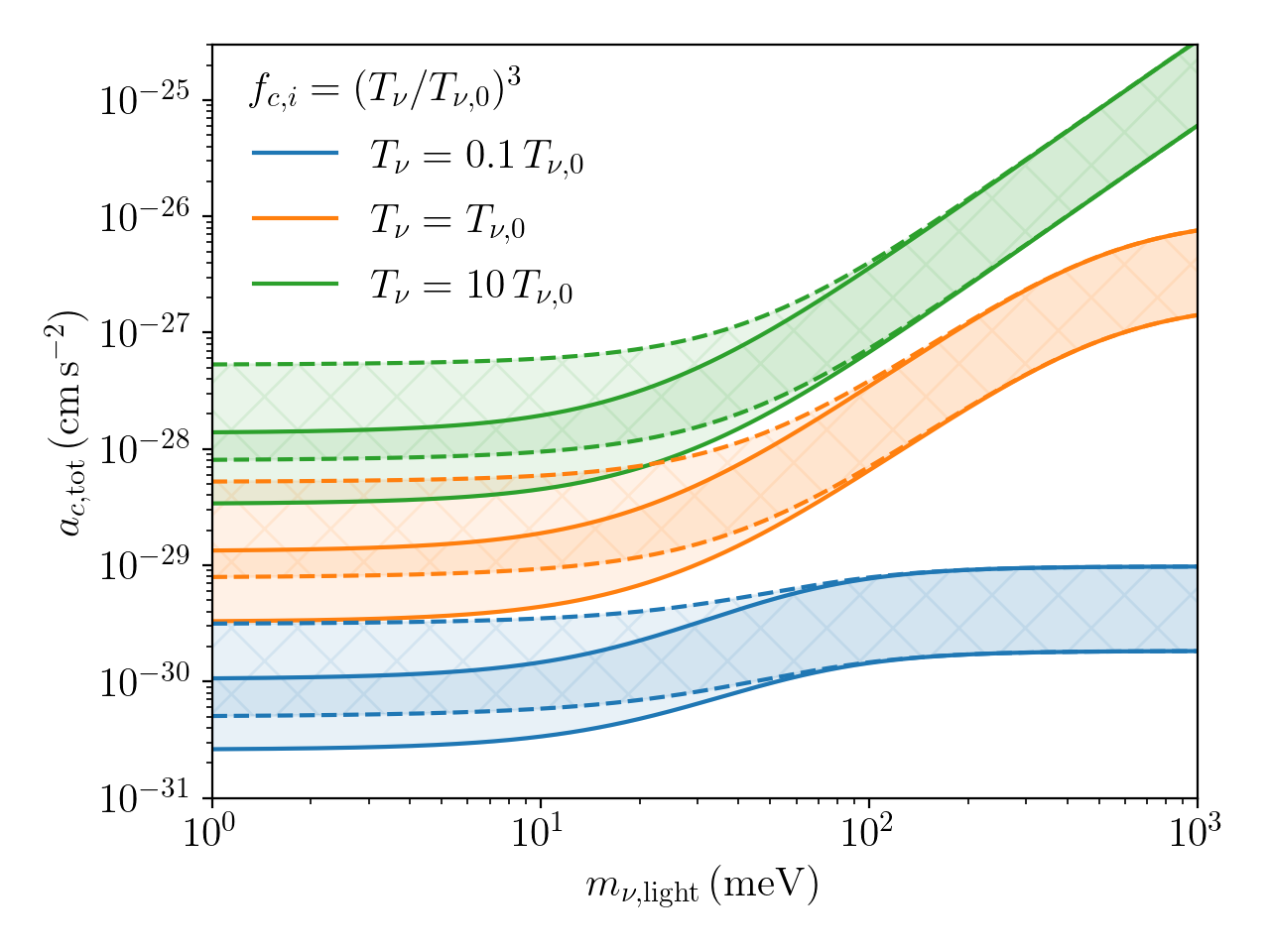}
\caption{\label{fig:accs} Total acceleration of a silicon target due to an unclustered neutrino wind as a function of the lightest neutrino mass, assuming the same temperature $T_\nu$ for all neutrino eigenstates. We plot using a solid contour and fill for NH and a dashed contour and hatch for IH, using the oscillation parameters from~\cite{Esteban:2020cvm}. Each shaded band corresponds to varying the efficiency of neutrino-electron bulk momentum transfer, $\varepsilon$, through $[0,1]$. Left: Assuming $n_{\nu_\alpha} = n_{\nu,0}$ for all temperatures. Right: Assuming equilibrium number density scaling, $n_{\nu_\alpha} \propto \,T_{\nu}^3$.}
\end{figure}

In practice, the size of this effect will depend on the properties of the material. For example, a target made of an non-metallic material where the electrons are strongly bound to their parent nuclei will recoil efficiently due to neutrino-electron scattering. On the other hand, a metal with delocalised electrons is susceptible to heating by neutrino-electron scattering, rather than transfer of momentum to the bulk solid. We note however that even in a good conductor, a large number of the electrons are still localised in core states. As such, it is reasonable to suggest that this heating effect will be small. Nevertheless, we choose to parameterise the total acceleration of a target due to an incident neutrino wind as
\begin{equation}
    a_{c,\mathrm{tot}} = a_{c,\mathrm{tot}}^N + \varepsilon\, a_{c,\mathrm{tot}}^e,
\end{equation}
where $a_{c,\mathrm{tot}}^e = \sum_\alpha a_c^e$ and $\varepsilon\in [0,1]$ parameterises the efficiency of momentum transfer from electrons to the bulk solid.

We show the total acceleration due to the coherent scattering of unclustered neutrinos on a silicon target in Figure~\ref{fig:accs}, considering a range of neutrino temperatures and including the contribution from neutrino-electron scattering. At low values of the lightest neutrino mass, the neutrino-induced accelerations are larger in the IH due to the larger masses of the two non-relativistic states $\nu_2$ and $\nu_3$. The distinction between NH and IH becomes less distinct at larger masses, when the masses of the three neutrino states become quasi-degenerate. In the left panel of Figure~\ref{fig:accs}, we additionally see that the acceleration becomes independent of temperature at large values of lightest neutrino mass, corresponding to the region where $T_\nu \ll T_{\nu,\mathrm{cold}}$.

At the KATRIN bound, $m_\nu = 0.8\,\mathrm{eV}$, assuming the NH we find a maximum acceleration $a_{c,\mathrm{tot}}=6.73\cdot 10^{-27}\,\mathrm{cm}\,\mathrm{s}^{-2}$ at $T_\nu = T_{\nu,0}$ in the absence of overdensities. We will use this as our benchmark value when discussing the sensitivity required to observe this effect in Section~\ref{sec:exp}.
    
\section{Experimental feasibility}\label{sec:exp}
There have been several proposals as to how we might possibly detect the tiny accelerations induced by a neutrino wind~\cite{Ringwald:2004np,Domcke:2017aqj,Smith:2003sy, Hagmann:1999kf}. The most widely discussed experimental setup uses a Cavendish-style torsion balance, which measures the torque induced by a difference in external forces acting on several connected test masses. This setup has historically been able to probe differential accelerations $\Delta a\simeq 10^{-13}\,\mathrm{cm}\,\mathrm{s}^{-2}$~\cite{Hagmann:1999kf}, and more recently as small as $\Delta a\simeq10^{-15}\,\mathrm{cm}\,\mathrm{s}^{-2}$ in tests of the weak equivalence principle~\cite{Wagner:2012ui}. Torsion balances utilising a test mass suspended by superconducting magnets have also been proposed~\cite{Hagmann:1999kf}, which would be sensitive to accelerations $\Delta a\simeq 10^{-23}\,\mathrm{cm}\,\mathrm{s}^{-2}$. These all fall short of the sensitivity required to observe this effect in the absence of significant overdensities, requiring a neutrino temperature $T_{\nu_\alpha} \simeq 15\,\mathrm{T_{\nu,0}}$ to be observable in the case that $f_{c,\alpha}=(T_{\nu_\alpha}/T_{\nu,0})^3$ for unclustered neutrinos. Equally, the torsion balance setup is unable to probe clustered relic neutrinos for masses $m_\nu < 0.8\,\mathrm{eV}$, where the clustering factor is expected to be $f_{c,i} \lesssim 50$~\cite{Zhang:2017ljh, Li:2011mw}.

One could also search for minute accelerations using the free-fall of test masses in low gravity environments. This technique has been applied to test the weak equivalence principle by the MICROSCOPE collaboration, who have been able to probe differential accelerations as small as $\Delta a\simeq3\cdot 10^{-14}\,\mathrm{cm}\,\mathrm{s}^{-2}$~\cite{Touboul:2017grn}. %albeit at frequencies $\Gamma \simeq 10^{-3}\,\mathrm{Hz}$.

\begin{figure}[tbp]
\centering 
\includegraphics[width=.49\textwidth]{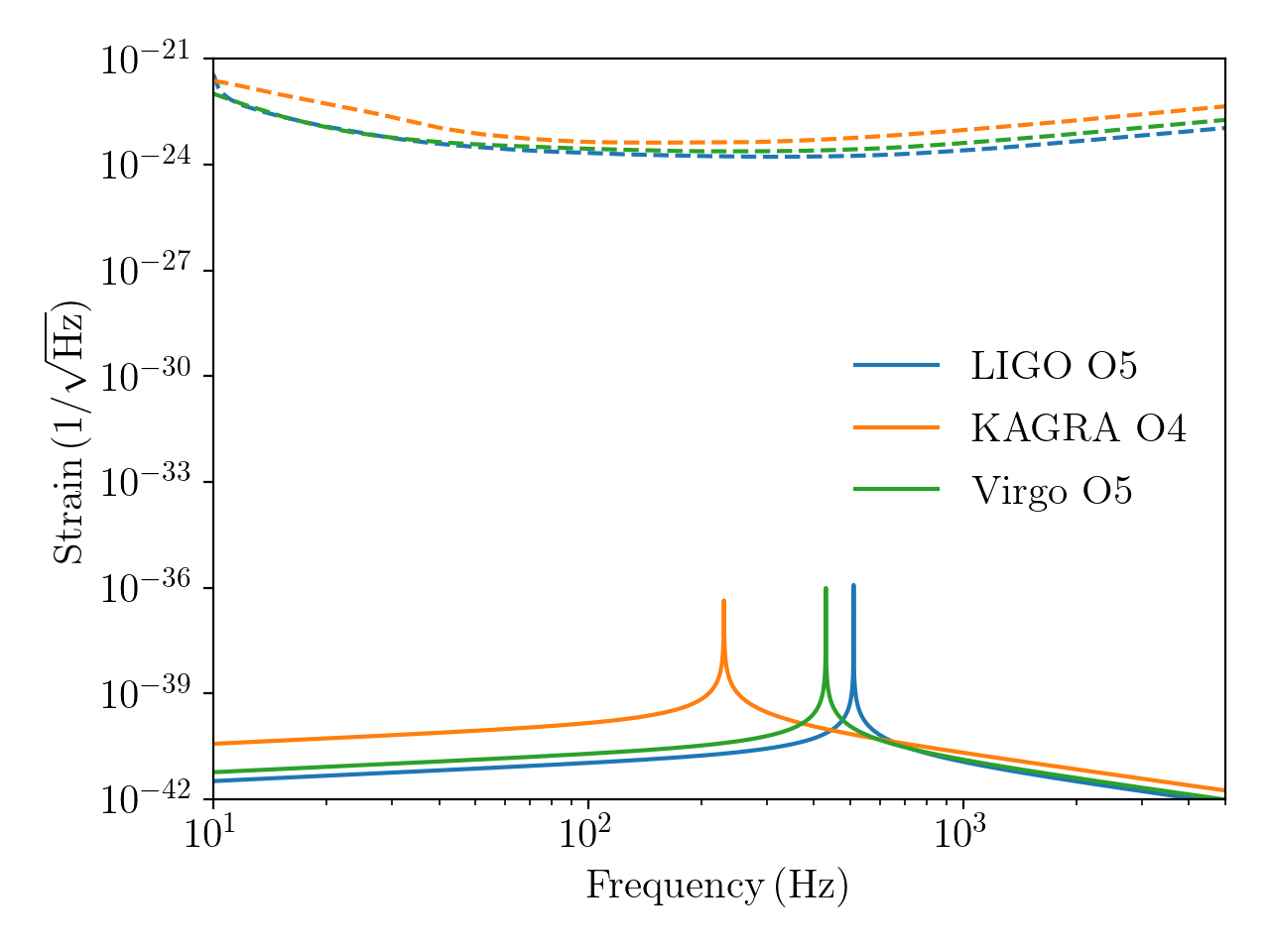}
\hfill
\includegraphics[width=.49\textwidth]{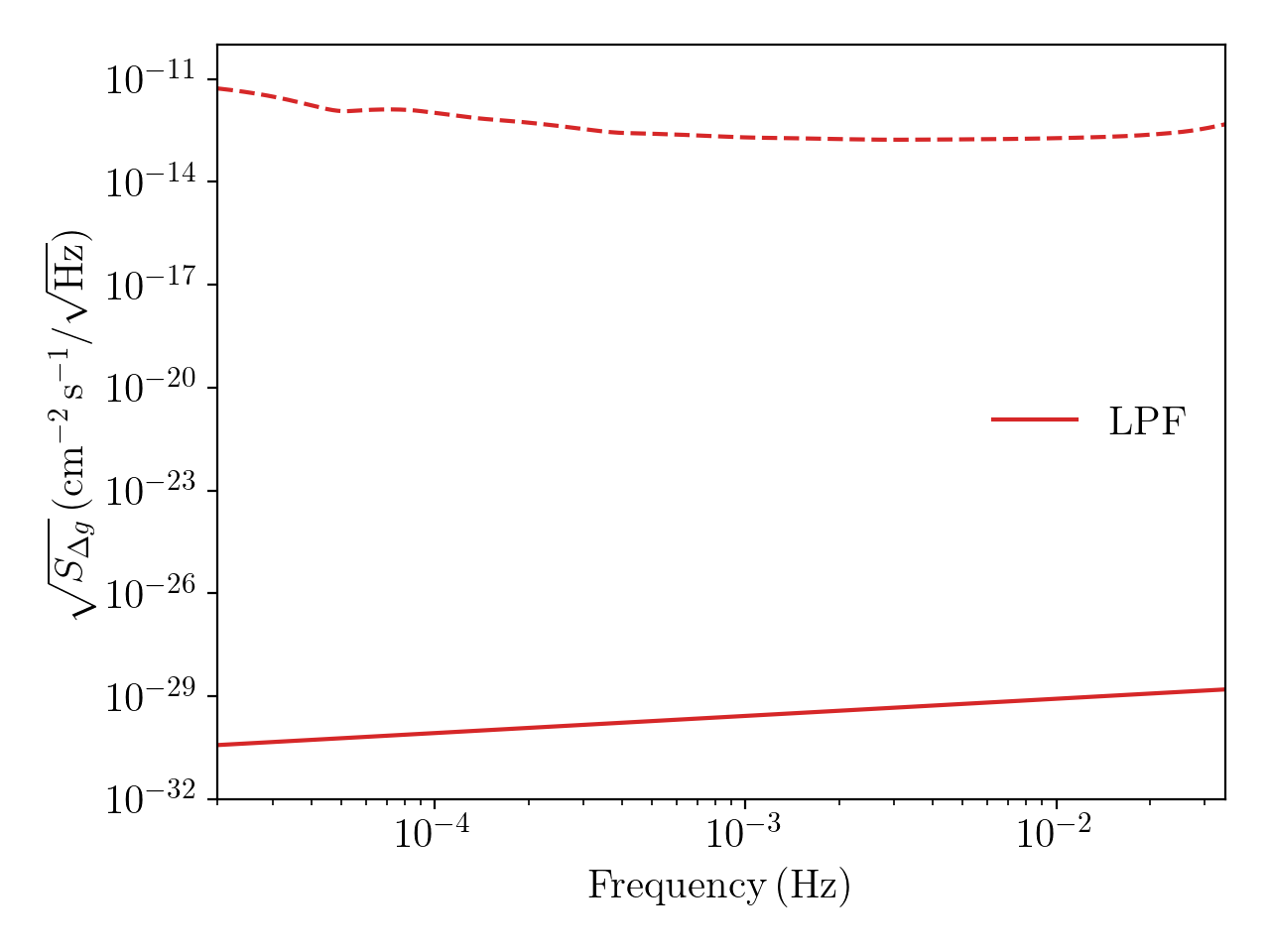}
\caption{\label{fig:GW} Left/right: Strain profiles/amplitude spectral density (solid lines) resulting from a single scattered neutrino in a gravitational wave detector, alongside the corresponding experiment sensitivity (dashed lines). For the experiments (LIGO, KAGRA, Virgo) we use the parameter values $f_r = (510,230, 430)\,\mathrm{Hz}$, corresponding to the first violin mode of the thermal suspension noise, $\xi = (8\cdot10^{-10},2.1\cdot 10^{-4} ,10^{-7})$, approximately equal to the half the suspension loss angles, $M = (40, 23, 42)\,\mathrm{kg}$ and $L = (4, 3, 3)\,\mathrm{km}$. We additionally $M = 2\,\mathrm{kg}$ for LISA Pathfinder (LPF). Sensitivities and other experimental parameters taken from several sources~\cite{KAGRA:2013pob, Armano:2018kix, LIGOScientific:2014pky,VIRGO:2014yos,Somiya:2011np} . We assume that the scattered neutrino transfers momentum $\Delta p_\nu = 1.6 \,\mathrm{meV}$, corresponding to $\bar\beta_\nu = 10^{-3}$ and $m_\nu = 0.8\,\mathrm{eV}$.}
\end{figure}

In an alternative proposal, Domcke and Spinrath~\cite{Domcke:2017aqj} suggested using a laser interferometer to measure the displacement of a pendulum from equilibrium as a result of an incident neutrino wind. Here the authors identified the signal frequency with the scattering frequency, found by inverting~\eqref{eq:accNN},
\begin{equation}\label{eq:scatteringFreq}
    \Gamma = \frac{a_{c,\mathrm{tot}}M}{\Delta p_\nu}\simeq 3.15\,\mathrm{kHz} \left(\frac{M}{40\,\mathrm{kg}}\right),
\end{equation}
where we have chosen the target mass to align with the $40\,\mathrm{kg}$ mirrors employed at LIGO~\cite{LIGOScientific:2016emj} and the far right hand side holds true for the reference scenario outlined at the end of Section~\ref{sec:gfsq}. 

The signatures of light dark matter scattering in gravitational wave detectors were later revised in~\cite{Lee:2020dcd} and found to be more akin to a series of instantaneous collisions with strain profiles peaking at frequencies not necessarily equal to the scattering frequency. As the collision frequency~\eqref{eq:scatteringFreq} is lower than the sampling rate of existing ground-based interferometers~\cite{LIGOScientific:2019hgc} and the average distance between scattered neutrinos far exceeds their wavelength, $\beta_\nu/\Gamma \gg \lambda_\nu$, we can treat each collision independently. In this case, the strain profile from a single scattering event\footnote{We assume here that the target is a single oscillator. In practice, gravitational wave detectors use a set of coupled harmonic oscillators. We further assume just a single resonance frequency, but note that an oscillator may have several, e.g the harmonics of the first violin mode in Figure~\ref{fig:GW}. See~\cite{Lee:2020dcd} and \cite{Saulson:1990jc} for comprehensive reviews.} is~\cite{Lee:2020dcd}
\begin{equation}
    h(f) = \frac{\sqrt{f}}{2\pi^2}\frac{\Delta p_\nu}{M L}\left|\frac{1}{f_r^2 - (f - if_r \xi)^2}\right|,
\end{equation}
where $L$ is the interferometer arm length, $f$ is the signal frequency, $f_r$ is the resonance frequency of the system $\xi\ll 1$ is related the damping of the oscillator. The situation differs somewhat for space-based interferometers where the masses are in free fall, making it difficult to discuss in terms of a harmonic oscillator. Instead, we measure the amplitude spectral density (ASD) of the differential acceleration between the test mass and the spacecraft, as suggested in~\cite{Lee:2020dcd},
\begin{equation}
    \sqrt{S_{\Delta g}(f)} = 2\sqrt{f}\frac{\Delta p_\nu}{M}.
\end{equation}
In Figure~\ref{fig:GW} we show the expected strain profiles for coherent neutrino scattering in the LIGO, KAGRA~\cite{Somiya:2011np} and Virgo~\cite{VIRGO:2014yos} terrestrial gravitational wave detectors, as well as the ASD for LISA Pathfinder (LPF)~\cite{Armano:2009zz}, the precursor to the space-based interferometer LISA~\cite{LISA:2017pwj}. Clearly, the strains expected from coherent neutrino scattering in gravitational wave detectors are far too small to be observed by any existing or proposed experiments. We further note that this is made worse for ground based interferometers by the fact that peaks in the thermal noise coincide with those of neutrino scattering. This method still has promise for the detection of dark matter, however, where the strain profiles for dark matter of mass $m_\mathrm{DM}$ are enhanced by a factor $\sim m_{\mathrm{DM}}/m_\nu$. 

\section{Conclusions}\label{sec:conc}
Observing relic neutrinos is an extraordinary challenge due to their low energy and tiny interaction cross sections, which leaves few avenues for their detection. In this paper, we have revisited the idea of detecting relic neutrinos using the small accelerations that they impart to a test mass via coherent scattering. Our main finding is a different scaling for the momentum transferred to the target by each scattering event to that in existing literature. This results in a faster growing acceleration with the neutrino mass at intermediate C$\nu$B temperatures. In addition, we have explored the neutrino-electron scattering contribution to the acceleration and shown that in the best case scenario it can exceed the neutrino-nucleus scattering component by almost an order of magnitude. We have also demonstrated that there exists an upper limit to the neutrino-induced acceleration which is independent of the neutrino mass and temperature, up to an overdensity factor. 

Despite the enhancement due to the electron contribution, we find that the accelerations are far too small to be observed by any existing or currently proposed experiment. However, a torsion balance utilising a superconductor-suspended test mass could detect the C$\nu$B in the presence of significant overdensities due to neutrino clustering or a higher neutrino temperature. We have also looked at the possibility of observing coherent relic neutrino scattering at gravitational wave experiments, and found that all existing and proposed experiments lack the required sensitivity to detect the tiny strains induced. 

\acknowledgments

I would like to thank Martin Spinrath for very helpful comments on the expected signals in gravitational wave detectors. I would additionally like to thank Chris Woodgate for a helpful discussion regarding the electronic structure of solids. This work is funded by an STFC studentship under the STFC training grant ST/T506047/1.


\end{document}